\documentclass[physrev,twocolumn,superscriptaddress]{revtex4-2}

\usepackage{graphicx}
\usepackage{amsmath,amssymb}
\usepackage{dcolumn}
\usepackage{bm}
\usepackage{color}
\usepackage{hyperref}
\hypersetup{colorlinks=true,breaklinks,linkcolor=blue,urlcolor=blue,citecolor=blue}
\usepackage{xspace}
\usepackage{multirow}

\newcommand{\vecp}{\bm{p}}
\newcommand{\vecr}{\bm{r}}

\newcommand{\vecx}{\bm{x}}
\newcommand{\vecy}{\bm{y}}
\newcommand{\vecz}{\bm{z}}
\newcommand{\vecu}{\bm{u}}
\newcommand{\vecrho}{\bm{\rho}}

\newcommand{\pF}{p^{\vphantom{\dagger}}_\mathrm{F}}

\newcommand{\dd}{{\mathrm{d}}}
\newcommand{\PHDG}{{\vphantom{\dagger}}}

\begin{document}

\title{Compressed Sensing of Compton Profiles for Fermi Surface Reconstruction:\\ Concept and Implementation}

\author{Junya Otsuki}
\affiliation{Research Institute for Interdisciplinary Science, Okayama University, Okayama 700-8530, Japan}
\author{Kazuyoshi Yoshimi}
\affiliation{Institute for Solid State Physics, University of Tokyo, Chiba 277-8581, Japan}
\author{Yoshinori Nakanishi-Ohno}
\affiliation{Faculty of Culture and Information Science, Doshisha University, Tatara Miyakodani 1-3, Kyotanabe, Kyoto 610-0394, Japan}
\author{Michael Sekania}
\affiliation{Institut f\"ur Physik, Martin-Luther Universit\"at Halle-Wittenberg, 06120 Halle/Saale, Germany}
\affiliation{Theoretical Physics III, Center for Electronic Correlations and Magnetism, Institute of Physics, University of Augsburg, 86135 Augsburg, Germany}
\affiliation{Center for Condensed Matter Theory and Quantum Computations, Ilia State University, 0162, Tbilisi, Georgia}
\author{Liviu Chioncel}
\affiliation{Theoretical Physics III, Center for Electronic Correlations and Magnetism, Institute of Physics, University of Augsburg, 86135 Augsburg, Germany}
\affiliation{Augsburg Center for Innovative Technologies, University of Augsburg, 86135 Augsburg, Germany}
\author{Masaichiro Mizumaki}
\affiliation{Japan Synchrotron Radiation Research Institute (JASRI), Sayo, Hyogo 679-5198, Japan}
\affiliation{Graduate School of Natural and Science, Okayama University, Okayama 700-8530, Japan}

\date{\today}

\begin{abstract}
Compton scattering is a well-established technique that can provide detailed information about electronic states
in solids.
Making use of the principle of tomography, it is possible to determine the Fermi surface from sets of Compton-scattering data with different scattering axes.
Practical applications, however, are limited due to long acquisition time required for measuring along enough number of scattering directions.
Here, we propose to overcome this difficulty using compressed sensing.
Taking advantage of a hidden sparsity in the momentum distribution, we are able to reconstruct the three-dimensional momentum distribution of bcc-Li, and identify the Fermi surface with as little as 14 directions of scattering data with unprecedented accuracy.
This compressed-sensing approach will permit further wider applications of the Compton scattering experiments.
\end{abstract}

\maketitle

\section{Introduction}
\label{sec:introduction}
The Compton scattering comprises the collision events in which photons (usually X-rays) are inelastically scattered by electrons in materials.
Since these electrons are in motion, the scattered radiation is Doppler-broadened and its measurement provides information on the electron momentum density (EMD) projected along the scattering direction~\cite{Cooper1971, Cooper2004}.
Compton scattering measurements
play an important role in investigations of the finite-temperature electronic structure, and supplies complementary information to other experiments such as the angle-resolved photoemission
spectroscopy (ARPES) and the de Haas--van Alphen measurement.

Experimental
Compton scattering studies on elemental Li and Al revealed marked influences of electronic correlations in particular in Li~\cite{Sakurai1995,Tanaka2001,Ohata2000}.
Recent applications to strongly correlated superconductors unveiled
the Fermi surface in 
cuprates La$_{2-x}$Sr$_x$CuO$_4$~\cite{Sakurai2011},
cobalt oxides Na$_x$CoO$_2$~\cite{Laverock2007}, and
doped iron-arsenides~\cite{Utfeld2010}.
Other applications to topical compounds include the study of the EMD around Dirac cones in graphene~\cite{Hiraoka2017},
the Fermi surface change across the metal-insulator transition in Ba$_{1-x}$K$_x$BiO$_3$~\cite{Hiraoka2005}, 
observation of a smeared Fermi surface in high-entropy alloys~\cite{Dugdale2006,Robarts2020},
and the temperature evolution between small and large Fermi surfaces in heavy-fermion compound YbRh$_2$Si$_2$~\cite{Guttler2021}.
Furthermore, magnetic Compton scattering using circularly polarized X-rays clarified the spin-dependent EMD of ferromagnetic iron and nickel~\cite{Kubo1990,Tanaka1993,Kakutani2003,ce.we.16,ja.se.21} and the orbital resolved occupations in Mn compounds~\cite{Koizumi2001}.
On the theoretical side, recent developments take account of electronic correlations by $GW$ approximation~\cite{Olevano2012}
and by the dynamical mean-field theory, which is applied to iron, nickel, and their alloy~\cite{Benea2012,Chioncel2014,be.mi.18}.
More recent proposals include an unexpected universal scaling predicted for the Compton profiles of alkali metals~\cite{Sekania2018},
and the detection of magnetoelectric multipoles
through the Compton scattering~\cite{Bhowal2021}.

\begin{figure}
    \centering
    \includegraphics[width=\linewidth]{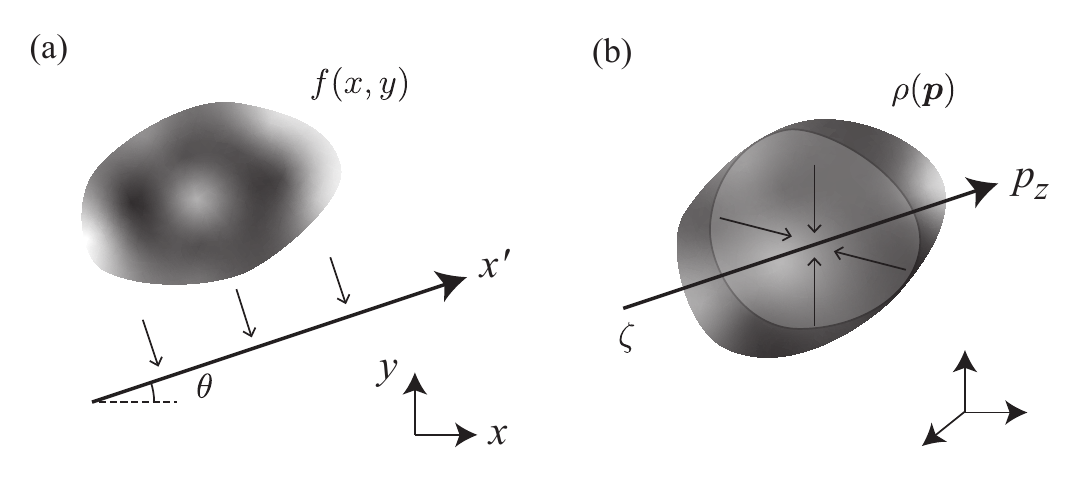}
    \caption{Schematics of (a) the CT for reconstruction of a two-dimensional function $f(x, y)$ and (b) the Compton profile represented as a double integral of three-dimensional function $\rho(\vecp)$.}
    \label{fig:CT_and_compton}
\end{figure}

The Compton scattering experiment measures the double differential cross section ${\dd^2\sigma/\dd\omega \dd\Omega}$.
Within the so-called impulse approximation~\cite{ch.wi.52,cu.de.71}, the cross section yields the Compton profile $J_{\zeta}(p_z)$, which is related to EMD, $\rho(\vecp)$, by a double integral~\cite{Cooper1971,Cooper2004}
\begin{align}
    J_{\zeta}(p_z) = \iint \rho(\vecp)\, \dd p_x \dd p_y.
    \label{eq:compton_profile}
\end{align}
Here, $p_z$ is chosen to be parallel to the scattering direction denoted by $\zeta$~[Fig.~\ref{fig:CT_and_compton}(b)].
Although the Compton profile $J_{\zeta}(p_z)$ possesses the information of $\rho(\vecp)$, the double integral
obscures
characteristics in $\rho(\vecp)$.
In particular, discontinuities in $\rho(\vecp)$ show up only as cusps in $J_{\zeta}(p_z)$, and hence the Fermi-surface features are difficult to identify from the experimental $J_{\zeta}(p_z)$ data.
Therefore, there is a need to improve the reconstruction of $\rho(\vecp)$ to enhance the capability of Compton scattering experiments to address open questions for fermiology.

The inverse problem of Eq.~\eqref{eq:compton_profile} can be regarded as a three-dimenensional extension of the computed tomography (CT).
Let $f(x, y)$ be a function in $x$-$y$ plane, and its one-dimensional projection 
$
    g_{\theta}(x') = \int \dd y'\, f(x, y)
$
is given,
where ${(x', y')}$ are the coordinate rotated from ${(x, y)}$ by angle $\theta$~[Fig.~\ref{fig:CT_and_compton}(a)].
Then, the original function $f(x, y)$ can be reconstructed,
provided that a set of $g_{\theta}(x')$ is available for a sufficiently dense distribution of $\theta$.
This principle has been applied to the Compton scattering to reconstruct the three-dimensional function $\rho(\vecp)$ from the Compton profile~[Fig.~\ref{fig:CT_and_compton}(b)]~\cite{Mijnarends1967,Mijnarends1969,Tanaka2001}.

Compared with the original CT,
the reconstruction of $\rho(\vecp)$
involves practical difficulties for following two reasons.
Firstly, the reconstruction of $\rho(\vecp)$ requires recovering of two axes eliminated by the double integral in Eq.~\eqref{eq:compton_profile}, while the original CT recovers only one axis.
Secondly, \emph{the number of the measurement axes $\zeta$ is limited} (about 10) for experimental reasons, while the angle $\theta$ in the CT is practically continuous.
Because of these difficulties, recent applications recover only one axis and employ the two-dimensional EMD projected onto a plane (e.g., $p_x$-$p_y$ plane)~\cite{Kontrym-Sznajd2003,Kontrym-Sznajd2004,Ketels2021}. This works for investigation of two-dimensional materials, in which the projected EMD still capture the feature of the Fermi surface. However, in order to investigate materials having a three-dimensional Fermi surface, full reconstruction of the three-dimensional EMD, $\rho(\vecp)$, is indispensable.

Following recent development in data-science techniques,
we are now able to improve the inversion process.
In this paper, we propose a method using compressed sensing to reconstruct the three-dimensional EMD.
Compressed sensing, first applied to MRI, is a data-processing technique that reduces required measurement data for obtaining a certain given precision of the density map~\cite{Candes2006,Candes2006a,Donoho2006,Lustig2007,Lustig2008}.
The key idea is that the final image of the density map is compressible, and the information, hence the number of measured Fourier signals, can be less than the number of pixels in the final image.
The success of the compressed sensing indicates that using characteristics of the EMD, there is a chance to carry out the Fermi-surface reconstruction with a much fewer number of scattering axes than it was required so far.

This paper is organized as follows.
We first review the concept of the compressed sensing in Section~\ref{sec:cs}.
Our reconstruction method and technical details in practical calculations are presented in Section~\ref{sec:method} and \ref{sec:procedure}, respectively.
Demonstrative results are presented in Section~\ref{sec:results} focusing on the noise of the input data.
The paper is summarized in Section~\ref{sec:summary}.

\section{Compressed sensing}
\label{sec:cs}

In this section, we review the fundamentals of the compressed sensing~\cite{Candes2008,Elad2010,Eldar2012,Krzakala2012,Otsuki2020} as a preliminary to its application to Compton profiles.
We consider a situation where a set of experimental data $\vecy$ is related to a physical quantity of interest, $\vecx$, by a linear equation ${\vecy=A \vecx}$.
Here, the sizes of vectors $\vecy$ and $\vecx$ are $M$ and $N$, respectively, and $A$ is an ${(M \times N)}$ matrix.
If $M<N$, namely, if the number of equations is less than the number of unknown variables, a solution for $\vecx$ is not uniquely determined (underdetermined systems).
Experimental errors further expand the set of possible solutions that satisfy ${\vecy=A \vecx}$ within error bars.
Finding physical solutions for $\vecx$ thus involves practical difficulties in realistic applications. 

Compressed sensing solves ${\vecy=A \vecx}$ for $\vecx$, assuming \emph{sparsity} in the solution $\vecx^{\ast}$.
This can be carried out by solving the optimization problem called generalized least absolute shrinkage and selection operator (LASSO)~\cite{Tibshiran1996,Tibshirani2011}.
In this case, the function $\mathcal{F}(\vecx)$ to minimize is given by
\begin{align}
    \mathcal{F}(\vecx) = \frac{1}{2} \| \vecy - A \vecx \|_2^2 + \lambda \| B \vecx \|_1^\PHDG,
    \label{eq:genlasso}
\end{align}
where $B$ is a non-square matrix,
and $\| \cdot \|_{\gamma}^\PHDG$ represents the $L_{\gamma}$ norm defined by
\begin{align}
    \| \vecx \|_{\gamma}^\PHDG &= \left( \sum_i |x_i|^{\gamma} \right)^{1/\gamma}.
\end{align}
The first term in Eq.~\eqref{eq:genlasso} yields a least-square fitting,
whereas the second term imposes a penalty for the absolute value of each component of $B\vecx$.
This penalty imposes solutions to have more zeros in $B\vecx$. 
A selected solution $\vecx^{\ast}$, thus, acquires sparsity in its linear combination $B\vecx^{\ast}$.

Clearly, the choice of the matrix $B$ is essential in the generalized LASSO.
Applications to MRI take advantage of the sparsity in the spatial variations of an expected image~\cite{Candes2006,Candes2006a,Donoho2006,Lustig2007, Lustig2008}.
The matrix $B$ in this case describes differences of intensities between neighboring pixels, which is called total variation~\cite{Rudin1992}.
With the aid of LASSO, measurement time required to obtain a certain resolution in the final result has shown to be reduced considerably.
The compressed sensing based on the $L_1$-norm regularization has been widely applied to various measurements~\cite{Moravec2007,Newton2012,Honma2014,Nakanishi-Ohno2016,Matsushita2016,Akai2018,Miyama2018,Tanaka2019,Yokoyama2019} and even to theoretical calculations~\cite{Nelson2013,Nelson2013a,Zhou2014,Seko2014,Tadano2015,Otsuki2017,Yoshimi2019}.

The regularization parameter $\lambda$ plays a crucial role in LASSO.
How to determine the optimal value of $\lambda$ will be demonstrated using explicit data in Section~\ref{sec:procedure}.

Finally, a comment on the matrix $A$ is in order.
For successful applications of compressed sensing, $A$ should be a \emph{dense} matrix as discussed below.
If $A$ is not dense, the matrix $A$ connects an element of $\vecx$ to only a few elements of input $\vecy$.
Hence, if some of these elements of $\vecy$ are missing, the corresponding element of $\vecx$ cannot be reproduced with accuracy. This might lead to a complete failure of the procedure, resulting in entirely wrong solution of $\vecx$.
If $A$ is a dense matrix, on the other hand, the lack of knowledge of some elements of $\vecy$, has only a diffuse effect over all elements of $\vecx$ and might lead to only minor errors in the outcome.
Moreover, a dense $A$ will cause a large number of degeneracies of possible solutions of $\vecx$, and the $L_1$-norm regularization will effectively work in choosing a sparse solution.

\section{Reconstruction Method}
\label{sec:method}

\subsection{Overview of reconstruction methods}

The Radon transform and the equivalent inverse formula found by Cormack in early $'60$~\cite{Cormack1963,Cormack1964} are the seminal works which allowed the development of current CT.
Mijnarends applied the method of Cormack to the problem in the angular correlation of positron annihilation radiation~\cite{Mijnarends1967, Mijnarends1969}, which involves the same inversion problem as Eq.~(\ref{eq:compton_profile}).
He represented $J_{\zeta}(p_z)$ and $\rho(\vecp)$ in terms of the spherical harmonics, $J_{lm}(p)$ and $\rho_{lm}(p)$, respectively.
Equation~(\ref{eq:compton_profile}) then forms an integral equation consisting of $J_{lm}(p)$, which is represented around the scattering axis, and $\rho_{lm}(p)$, which is represented in the crystal coordinate.
This complicated equation has been solved analytically.
Therefore, once $J_{lm}(p)$ are obtained from experimental data, they are immediately converted into $\rho_{lm}(p)$, and thus $\rho(\vecp)$, using the analytical solution.
This method has also been applied to the reconstruction of two-dimensional projected EMD~\cite{Kontrym-Sznajd2003,Kontrym-Sznajd2004,Ketels2021}.

An alternative approach uses the Fourier transform as is now common in practical appliations of CT.
Tanaka~\textit{et al.} applied the direct Fourier transform method to the Compton profiles with elaborate consideration of the error propagation~\cite{Tanaka2001}.
They demonstrated reconstruction of the three-dimensional EMD from experimentally measured Compton profiles of a lithium metal.
There is a room for improvement in the fact that the truncation of the Fourier series results in artificial oscillations in the final result of $\rho(\vecp)$, which make it difficult to identify the discontinuity in $\rho(\vecp)$ (Fermi surface).

From the point of view of the compressed-sensing technique, the direct Fourier transform method is more suitable than the Cormack's method for the following reasons.
As described in Sec.~\ref{sec:cs}, successful applications of the compressed sensing require the transformation matrix $A$ to be \emph{dense}.
The Cormack's method is represented in the polar coordinate, in which different radial coordinates are decoupled. Therefore, the matrix $A$ is sparse.
In the direct Fourier transform method, on the other hand, the matrix $A$ corresponds to the Fourier basis $e^{i\vecp \cdot \vecr}$, in which each real-space component is represented with the whole Fourier components. Therefore, the matrix $A$ is dense and satisfies the requirement of the compressed sensing.

\subsection{Direct Fourier transform method}

\begin{figure}
    \centering
    \includegraphics[width=0.7\linewidth]{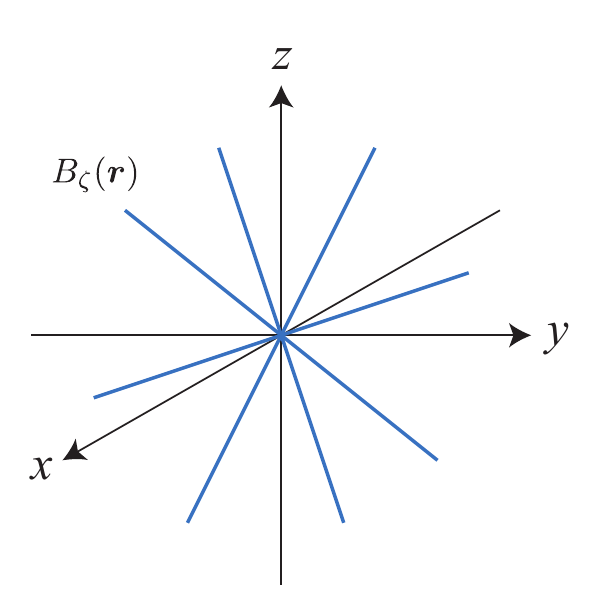}
    \caption{Schematic figure of the distribution of lines on which $B(\vecr)$ is obtained by the Fourier transform of the Compton profiles $J_{\zeta}(p_z)$ using Eq.~\eqref{eq:B_z}.}
    \label{fig:coordinate}
\end{figure}

We review the direct Fourier transform method by Tanaka \textit{et al.} in Ref.~\cite{Tanaka2001}.
We define the Fourier transform of the momentum density $\rho(\vecp)$ by $B(\vecr)$:
\begin{align}
    B(\vecr) = \iiint \rho(\vecp) e^{i\vecp \cdot \vecr}\, \dd\vecp.
    \label{eq:B_by_rho}
\end{align}
Substituting ${\vecr=(0,0,z)}$ in a coordinate system with $z$ axis being parallel to the scattering direction $\zeta$, we obtain 
\begin{align}
    B_{\zeta}(0, 0, z) = \int J_{\zeta}(p_z) e^{ip_z z} \, \dd p_z.
    \label{eq:B_z}
\end{align}
Here, we used Eq.~\eqref{eq:compton_profile} to replace $\rho(\vecp)$ with $J_{\zeta}(p_z)$.
The subscript $\zeta$ for $B$ is to indicate the direction of the $z$ axis.
Compton profiles $J_{\zeta}(p_z)$ measured on several scattering directions $\zeta$ yield $B(\vecr)$ on the corresponding lines in the real space as shown in Fig.~\ref{fig:coordinate}.
The inverse transformation of Eq.~\eqref{eq:B_by_rho} is given by
\begin{align}
    \rho(\vecp) = (2\pi)^{-3}\iiint B(\vecr) e^{-i\vecp \cdot \vecr} \, \dd\vecr.
    \label{eq:inverse_fourier}
\end{align}
In order to perform this integral, we need $B(\vecr)$ in the whole $\vecr$ space.
In Ref.~\cite{Tanaka2001}, $B(\vecr)$ obtained on several lines (Fig.~\ref{fig:coordinate}) is interpolated for arbitrary $\vecr$, and then the inverse transformation is carried out to reconstruct $\rho(\vecp)$.

\subsection{Application of compressed sensing}

In the inverse Fourier transform in Eq.~\eqref{eq:inverse_fourier}, missing information in $B(\vecr)$ was filled by interpolation, which could result in a reduction of accuracy.
In the following, we directly solve Eq.~\eqref{eq:B_by_rho} for $\rho(\vecp)$ without an interpolation by applying the compressed-sensing technique.

We represent the integral in Eq.~\eqref{eq:B_by_rho} with a discrete sum over $\vecp_j$ on a uniformly spaced grid, $\Delta p_\mathrm{calc}$, within a cube of volume $(2P_\mathrm{max})^3$.
The cube is taken to be sufficiently large so that the whole region where $\rho(\vecp_j)$ is finite is covered.
Then, Eq.~\eqref{eq:B_by_rho} is represented as
\begin{align}
    B_i = \sum_j A_{ij} \rho_j,
    \label{eq:B_by_rho-2}
\end{align}
where ${B_i \equiv B(\vecr_i)}$, ${\rho_j \equiv \rho(\vecp_j) \Delta p_\mathrm{calc}^3}$, and ${A_{ij} \equiv e^{i\vecp_j \cdot \vecr_i}}$.
$\rho_j$ is defined on a dense grid that covers the whole region, whereas $B_i$ is given only on lines that are computed from several Compton profiles in Eq.~\eqref{eq:B_z}.
Therefore, this linear equation forms an underdetermined system that has a fewer number of equations than the number of unknown variables.
Filling interpolated values in $B(\vecr)$ is a way to supply additional equations to make the system of linear equations solvable.

Instead of increasing the number of equations, we \emph{reduce} the number of variables that need to be determined.
To this end, we suppose that $\rho(\vecp)$ is constant, i.e., ${\nabla \rho(\vecp)=0}$, in an extensive region.
This is true away from the Fermi surface, where the energy bands are either fully occupied or empty.
Such a solution can be obtained by minimizing the following function of the form of the generalized LASSO (see Section~\ref{sec:cs}):
\begin{align}
  \begin{split}
    \mathcal{F}(\{ \rho_j \}) &=
    \frac{1}{2} \sum_{i \in \text{measured}} \left[ B_i - \sum_j A_{ij} \rho_j \right]^2
    \\
    &+ \lambda \sum_j \sum_{\xi=p_x, p_y, p_z} \left| \sum_{j'} (D_{\xi})_{jj'} \rho_{j'} \right|.
  \end{split}
    \label{eq:lasso-fourier}
\end{align}
Here, the summation in the first term is taken over $B_i$ computed from the measured Compton profiles.
$D_{\xi}$ is a matrix that represents the derivative ${\partial/\partial \xi}$. With the first-order forward difference, its explicit expression is given by
${(D_{\xi} \rho)_j = \rho_{j(+\xi)}-\rho_{j}}$, where the index ${j(+\xi)}$ denotes the coordinates one-point ahead of $\vecp_j$ to the direction $\xi$~\footnote{We omitted the factor ${1/\Delta p}$ because it only changes the scale of $\lambda$.}.
The second term in Eq.~\eqref{eq:lasso-fourier} forces the solution to have ${\partial \rho(\vecp) / \partial \xi=0}$, keeping the first term within a certain range.
To what extent the second term affects the solution
is controlled by the regularization parameter $\lambda$, which will be discussed in details in Section~\ref{sec:cv}.

There are two additional relations that $\rho(\vecp)$ should fulfill.
One is non-negativity
\begin{align}
    \rho(\vecp)\geq0,
\end{align}
and the other is the sum rule
\begin{align}
    \iiint \rho(\vecp) \, \dd \vecp = \int J_{\zeta} (p_z)\, \dd p_z \equiv n,
    \label{eq:sumrule}
\end{align}
which is obtained by integrating Eq.~\eqref{eq:compton_profile} over $p_z$.
Here, the value $n$ represents the total number of electrons in a unit cell.
In the discrete representation, the above two relations are written as
\begin{align}
    \rho_j \geq 0\,, \qquad \sum_j \rho_j = n,
    \label{eq:constraints}
\end{align}
Our goal is to minimize the function $\mathcal{F}$ in Eq.~\eqref{eq:lasso-fourier} with respect to $\rho_j$ under the constraints, Eq.~\eqref{eq:constraints}.

\section{Calculation procedure}
\label{sec:procedure}

\subsection{Compton profile data}

\begin{figure*}[tb]
    \centering
    \includegraphics[width=\linewidth]{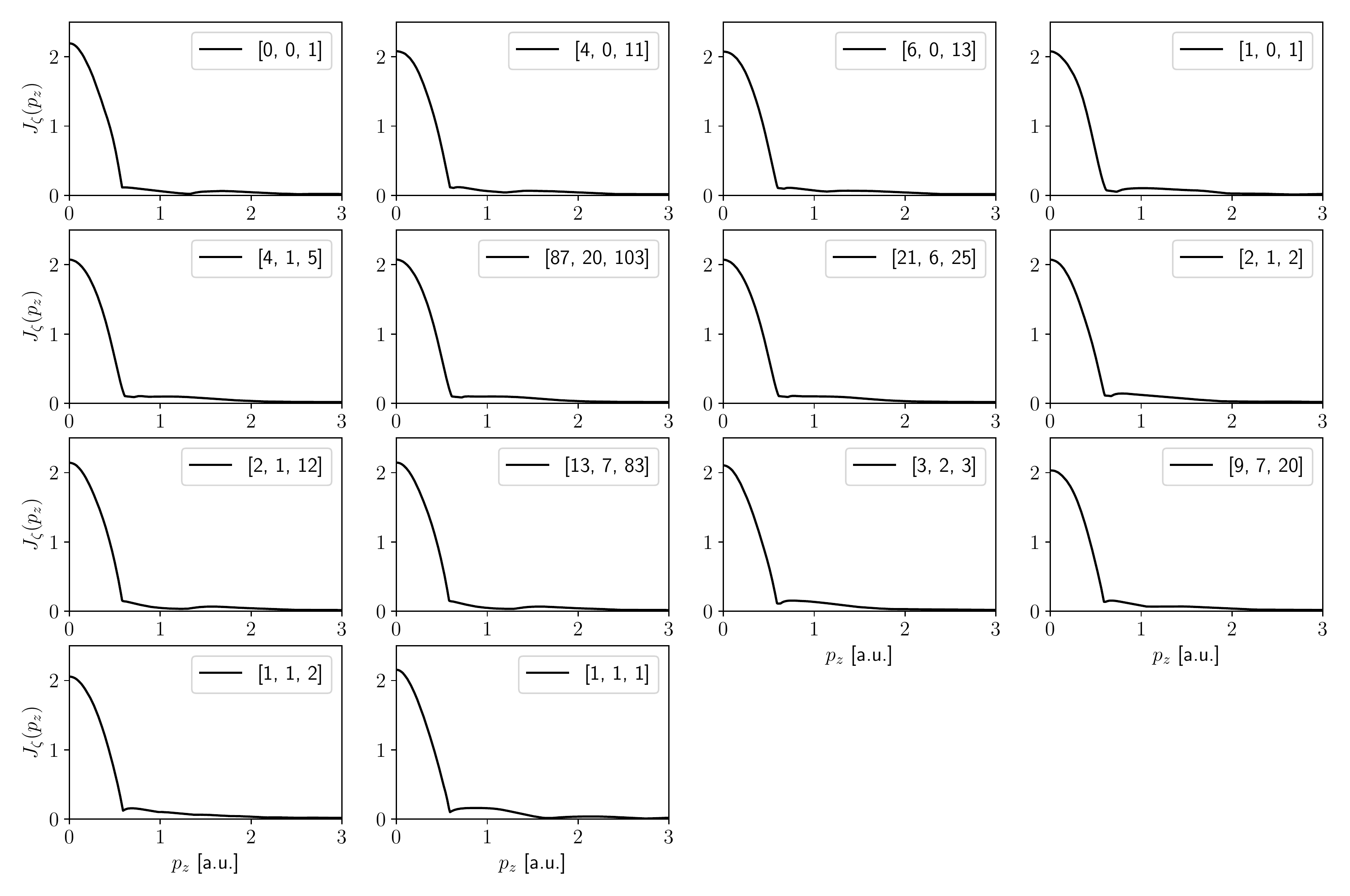}
    \caption{Compton profiles $J_{\zeta}(p_z)$ for Li metal computed on 14 axes.}
    \label{fig:Li_input}
\end{figure*}

To demonstrate the performance of our method described in the previous section, we apply it to bcc-Li, which has been addressed by the direct Fourier transform method~\cite{Tanaka2001}.
We prepare both the Compton profiles $J_{\zeta}(p_z)$ and the EMD $\rho(\vecp)$ by the first-principles calculations.
The reconstructed $\rho(\vecp)$ will be verified by comparing with $\rho(\vecp)$ directly computed without the reconstruction.

The electronic structure of alkali metals is calculated within density functional theory (DFT)~\cite{jo.gu.89,jone.15} using the spin-polarized relativistic Korringa-Kohn-Rostoker (SPR-KKR) method~\cite{eb.ko.11}.
With the local spin-density approximation (LSDA) for the exchange correlation potential~\cite{vo.wi.80}, the spin-resolved EMD are computed from the corresponding LSDA Green functions.
The self-consistent LSDA calculations are performed with a  ${62\times62\times62}$ mesh in the Brillouin zone~\cite{eb.ko.11}.
$J_{\zeta}(p_z)$ and $\rho(\vecp)$ are obtained by the energy integral in the complex plane on a semi-circular contour with $32$ points and a rectangular grid in the momentum space with a cutoff ${|\vecp|_\mathrm{max}=10}$\,a.u.~\cite{be.ma.06}.
The step size of the momentum is $0.01$\,a.u. for $J_{\zeta}(p_z)$ and $0.002$\,a.u. for $\rho(\vecp)$.
We normalize $J_{\zeta}(p_z)$ to satisfy the sumrule in Eq.~(\ref{eq:sumrule}).

Figure~\ref{fig:Li_input} shows $J_{\zeta}(p_z)$ computed for 14 directions.
The results for the principal directions, $[001]$, $[110]$, and $[111]$, have been published in Ref.~\cite{Sekania2018}.
The bin size of $J_{\zeta}(p_z)$ is ${\Delta p_\mathrm{exp} = 0.01}$\,a.u., which is comparable to the experimental one ${\Delta p_\mathrm{exp} = 0.02}$ in Ref.~\cite{Tanaka2001}.
To simulate experiments, we add Gaussian noise on $J_{\zeta}(p_z)$.
The width of the Gaussian distribution $\sigma$ is ${\sigma=10^{-1}}$, $10^{-2}$, or $10^{-3}$.
Specific features of $J_{\zeta}(p_z)$ are the parabola-like shape for ${p_z < \pF}$, first cusp at ${p_z = \pF}$, and the tails for ${p_z > \pF}$. 
The value of ${\pF = 0.58}$\,a.u. has been found from a precise computation using the enhanced momentum cutoff. 
Higher momentum contributions to $\rho(p)$ are frequently discussed (see Ref.~\cite{Cooper2004} and references therein)
and constitute a clear evidence for Umklapp processes.
Note also that the observed anisotropy of the Compton profiles is a consequence of the directional anisotropy of the bcc lattice.

\subsection{Fourier transform}

We first perform the Fourier transform of the Compton profiles $J_{\zeta}(p_z)$ in Eq.~\eqref{eq:B_z} to obtain its real-space representation $B_{\zeta}(0, 0, z)$.
Since $J_{\zeta}(p_z)$ is an even function, i.e., ${J_{\zeta}(p_z)=J_{\zeta}(-p_z)}$, the transform is represented as a discrete cosine transformation and $B(\vecr)$ is real.
The explicit expression for the discrete cosine transformation is presented in Appendix~\ref{app:fourier}.

\begin{figure}
    \centering
    \includegraphics[width=0.95\linewidth]{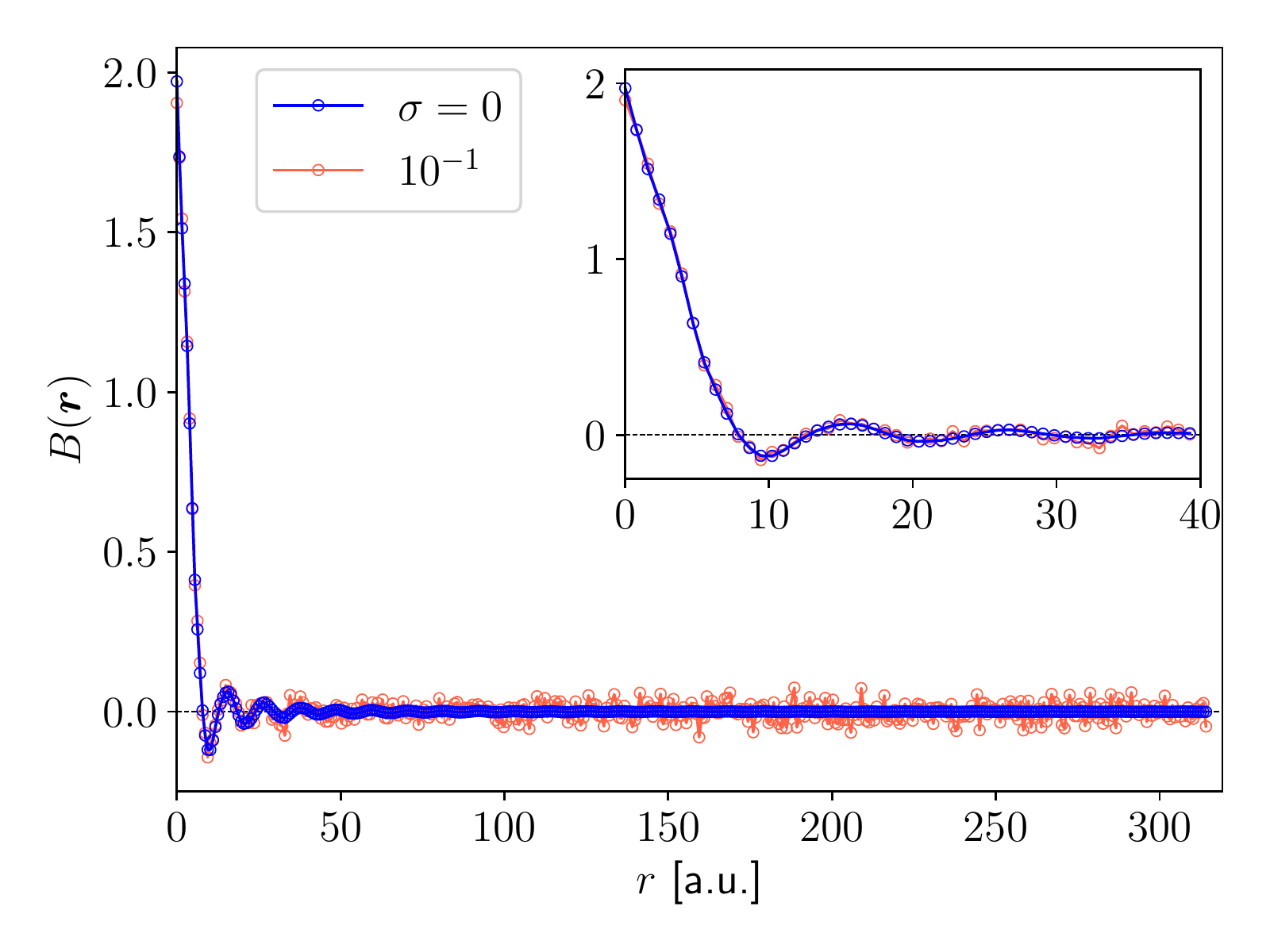}
    \caption{$B(\vecr)$ on the axis ${\zeta=[001]}$. The blue and red circles show results computed from $J_{\zeta}(p_z)$ without noise and with noise of ${\sigma=10^{-1}}$, respectively. The inset shows an enlarged view around ${r=0}$.}
    \label{fig:Br}
\end{figure}

Figure~\ref{fig:Br} shows $B(\vecr)$ along the axis ${\zeta=[001]}$.
The upper boundary of $r$ is given by ${r_\mathrm{max}=\pi/\Delta p_\mathrm{exp}} {\approx 314.2}$\,a.u.
The real-space step size $\Delta r$ is ${\Delta r=\pi/p_\mathrm{max}} {\approx 0.785}$\,a.u.
Figure~\ref{fig:Br} compares $B(\vecr)$ computed from $J_{\zeta}(p_z)$ with and without noise.
It is clear that the influence of noise is relatively large in the large-$r$ region, because $B(\vecr)$ decays with increasing $r$.

Although $B(\vecr)$ is obtained up to sufficiently large-$r$ region, we truncate these data before going to the next step for the reasons mentioned below.
As will be described in Section~\ref{sec:LASSO},
the momentum step size $\Delta p_\mathrm{calc}$ in the calculation of $\rho(\vecp)$ is limited because of the computer memory and the computation time for solving the LASSO optimization problem. Hence, in the ordinary situation,
$\Delta p_\mathrm{calc} > \Delta p_\mathrm{exp}$,
where $\Delta p_\mathrm{exp}$ is the bin size of the Compton profiles.
This results in a periodicity in $B(\vecr)$ evaluated from $\rho(\vecr)$ with the period ${\pi/\Delta p_\mathrm{calc}}$, which is smaller than the upper boundary $r_\mathrm{max}$ of the input data $B(\vecr)$.
Therefore, $B(\vecr)$ has to be truncated at ${\pi/\Delta p_\mathrm{calc} \equiv r_\mathrm{cutoff}}$.

\subsection{Solving LASSO}
\label{sec:LASSO}

The momentum points $\vecp_j$ for representing $\rho(\vecp)$ is constructed with a linear mesh between $-P_\mathrm{max}$ and $P_\mathrm{max}$ for each axis.
We set ${P_\mathrm{max}=3}$\,a.u. in this paper.
The number of points, $L$, for each axis is fixed at ${L=121}$.
The momentum step size is thus ${\Delta p_\mathrm{calc} = 0.05}$\,a.u.

We apply the symmetry operations against a set of ${N=L^3}$ momenta, $\{ \vecp_j \}$, to reduce the number of points.
The crystals of elemental alkali metals have $O_h$ point-group symmetry.
There are 48 symmetry operations and therefore only ${N_\mathrm{irr} \simeq N/48}$ momenta are inequivalent.
With this property, we can reduce the number of grid points.
For details of how to find equivalent points and how to integrate the symmetry features into the optimization problem in Eq.~\eqref{eq:lasso-fourier}, see Appendix~\ref{app:symmetry}.

For the irreducible set of momenta $\{ \tilde{\vecp}_j \}$, we solve the optimization problem in Eq.~\eqref{eq:lasso-fourier} under constraints, Eq.~\eqref{eq:constraints}.
We use the alternating direction method of multipliers (ADMM)~\cite{Boyd2010}, which is presented in details in Appendix~\ref{app:admm}.
The calculation time grows as $O(N_\mathrm{irr}^3)$, which limits the feasible maximum system size, and hence the resolution of the final image of $\rho(\vecp)$.
We thus chose $N=121^3$, which corresponds to $N_\mathrm{irr}=4.0 \times 10^3$.

\begin{figure}
    \centering
    \includegraphics[width=0.98\linewidth]{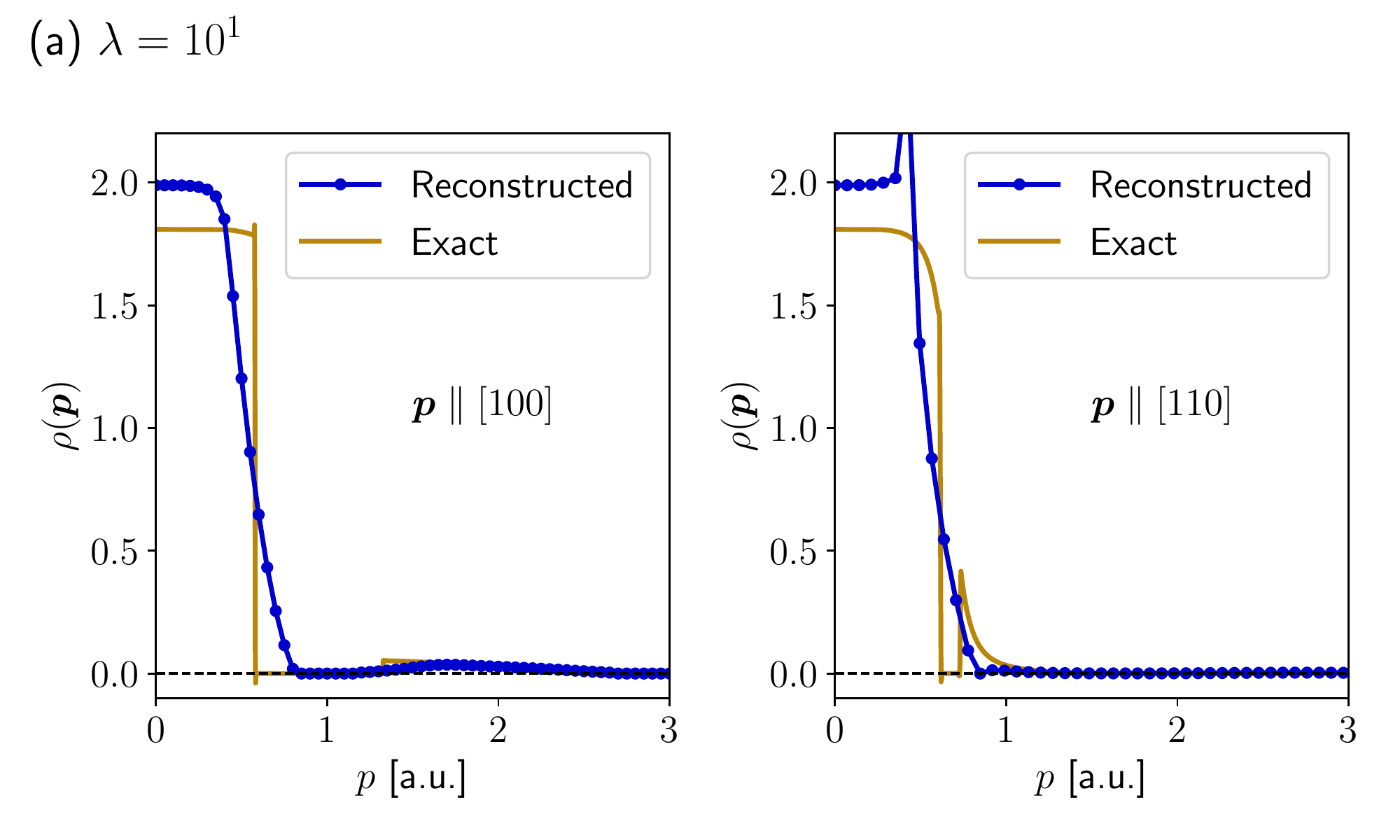}
    \includegraphics[width=0.98\linewidth]{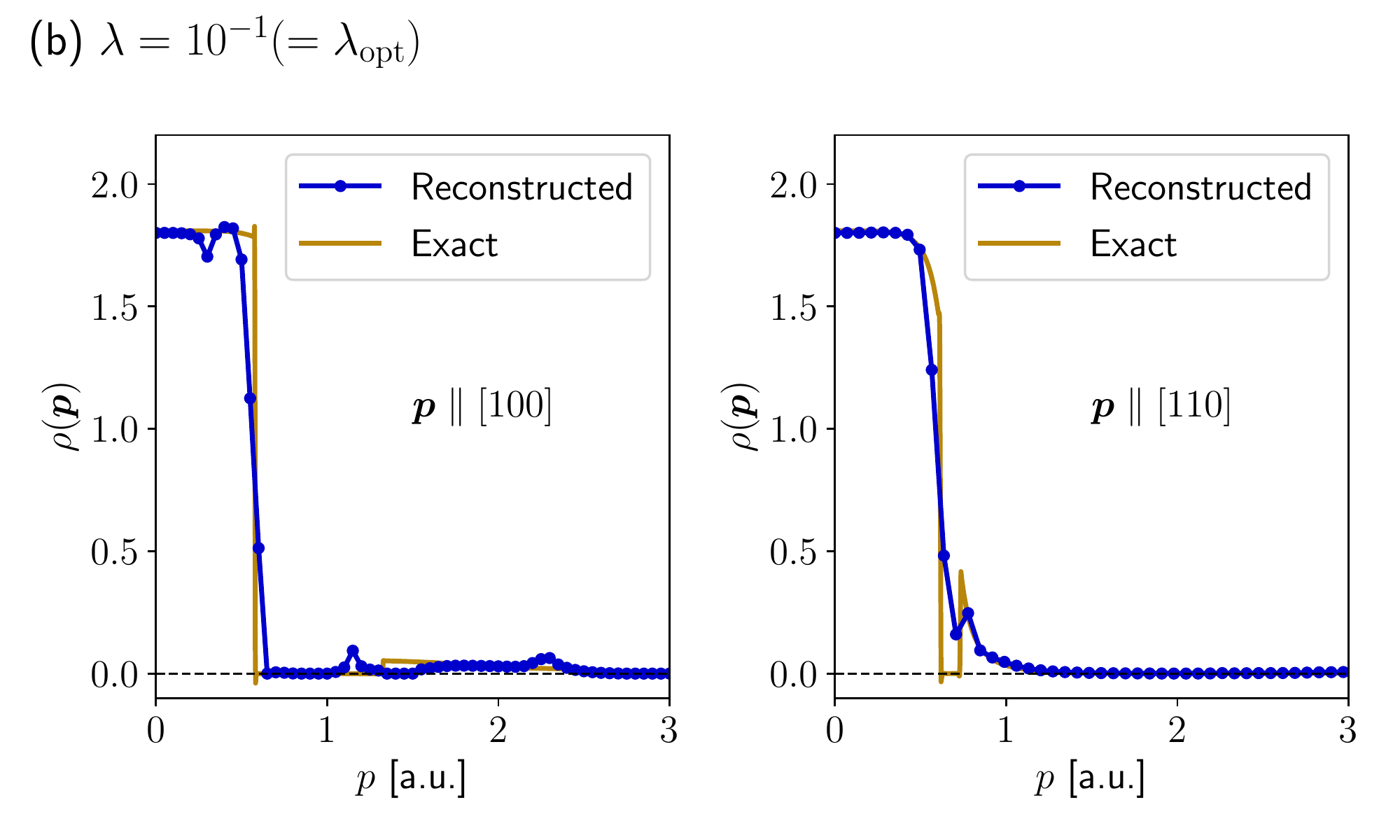}
    \includegraphics[width=0.98\linewidth]{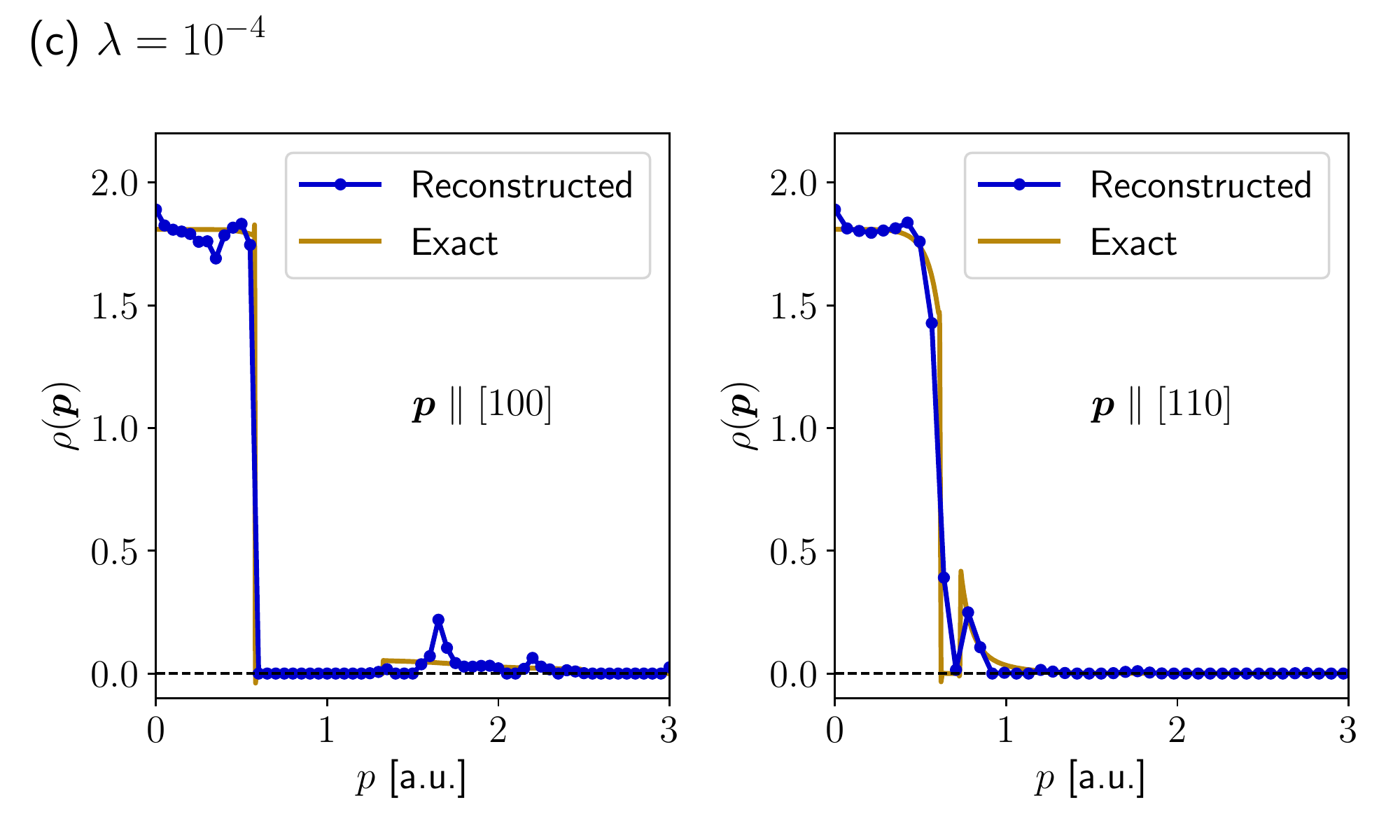}
    \caption{The reconstructed EMD $\rho(\vecp)$ along ${\vecp \parallel [100]}$ (Left panel) and ${\vecp \parallel [110]}$ (Right panel) compared with the exact one. The regularization parameter $\lambda$ is varied among (a) ${\lambda=10^{1}}$, (b) ${\lambda=10^{-1}}$ [the optimal value ${\lambda_\mathrm{opt}}$ determined by the cross validation method (see Section~\ref{sec:cv})], and (c) ${\lambda=10^{-4}}$.
    The noise is ${\sigma=10^{-3}}$.}
    \label{fig:np_example}
\end{figure}

Figure~\ref{fig:np_example} shows representative results for reconstructed $\rho(\vecp)$ along two symmetry axes, ${\vecp \parallel [100]}$ and $[110]$.
Results are shown for three values of the regularization parameter $\lambda$.
When $\lambda$ is large [Fig.~\ref{fig:np_example}(a)], $\rho(\vecp)$ tends to be flat except the region near the Fermi momentum ${\pF \approx 0.6}$\,a.u.. Furthermore, the discontinuity at ${p=\pF}$ is broadened.
In the opposite case with small $\lambda$ [Fig.~\ref{fig:np_example}(c)], 
we can identify the discontinuity at ${p=\pF}$ as well as some small features between ${p=1.3}$\,a.u. and ${p=2.5}$\,a.u. for ${\vecp \parallel [100]}$ and a structure around ${p=0.8}$\,a.u. for ${\vecp \parallel [110]}$.
But, there are some artificial features too, e.g., a hump at ${p=0}$ for ${\vecp \parallel [110]}$. 
Between the two limits, we can obtain a reasonable result that shows the physical structure well and exhibits less unphysical features [Fig.~\ref{fig:np_example}(b)].
We remark that our result does not show artificial oscillations as observed in the original direct Fourier transform method~\cite{Tanaka1993}.
This is due to the regularization term, which makes $\rho(\vecp)$ as flat as possible. Consequently, we achieve a clear discontinuity at the Fermi momentum without artificial oscillations.

\subsection{Determination of the regularization parameter}
\label{sec:cv}

\begin{figure}
    \centering
    \includegraphics[width=0.95\linewidth]{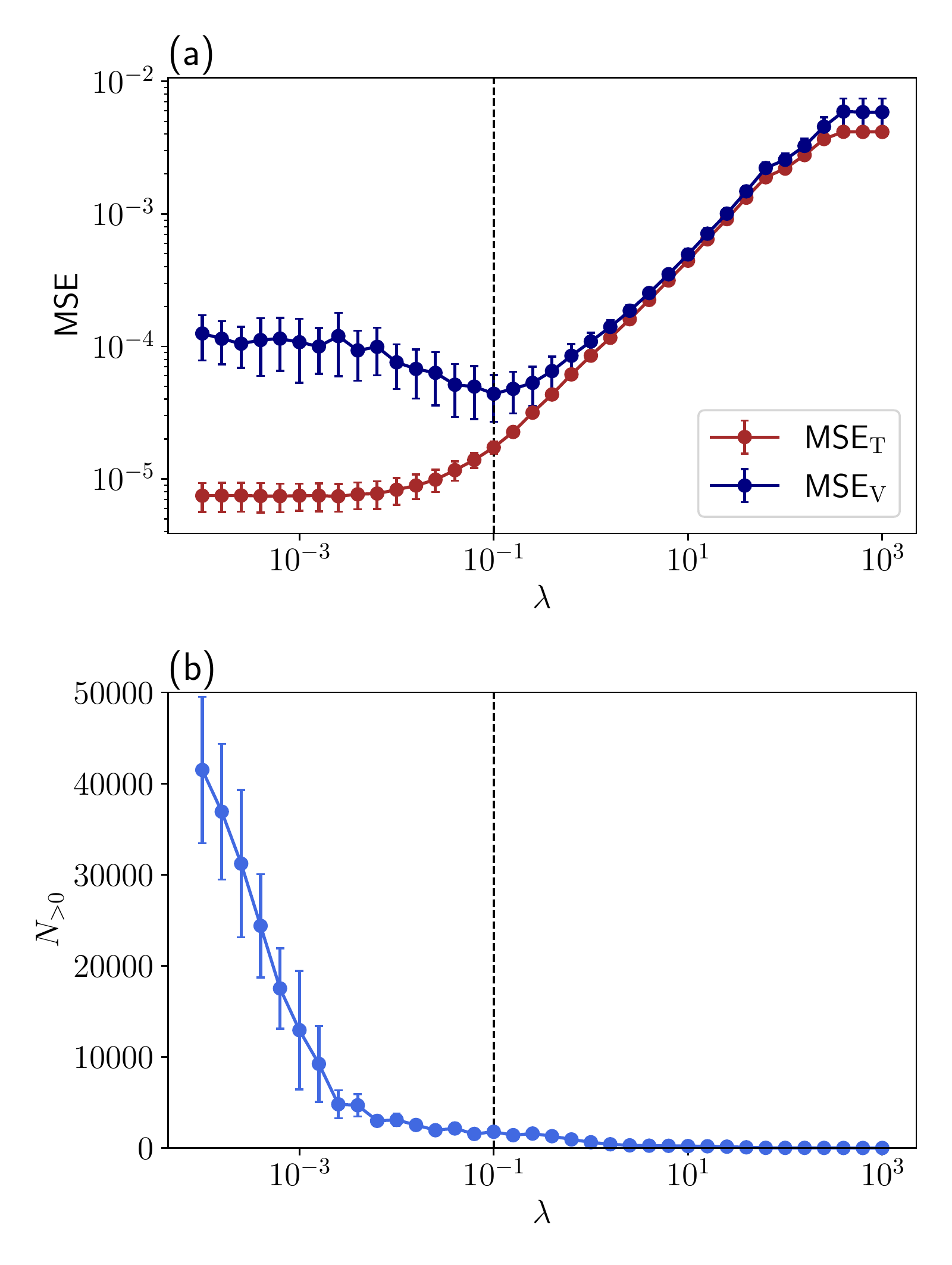}
    \caption{(a) The training error MSE$_\mathrm{T}$ and the validation error MSE$_\mathrm{V}$, and (b) the number of nonzero components $N_{>0}$ in the $L_1$-norm regularization term as a function of the regularization parameter $\lambda$. The dashed vertical line indicates the optimal $\lambda$, where MSE$_\mathrm{V}$ is minimized.}
    \label{fig:cv}
\end{figure}

For determining an optimal value of $\lambda$ in an unbiased manner, we use the cross validation (CV) method.
An application to LASSO is presented, for example, in Ref.~\cite{Otsuki2020}.

The input data $\vecy$ is split into $K$ groups randomly.
$K$ is fixed at ${K=5}$ in this paper.
${(K-1)}$-groups of data is denoted by $\vecy^\PHDG_\mathrm{T}$, and the rest one group of data is denoted by $\vecy^\PHDG_\mathrm{V}$. Here, the subscripts T and V stand for training and validation, respectively.
The LASSO is solved with $\vecy^\PHDG_\mathrm{T}$ as an input.
More precisely, the summation of the index $i$ in Eq.~\eqref{eq:lasso-fourier} is taken over the subset $\vecy_\mathrm{T}$. 
The converged solution $\vecx^{\ast}$ is validated with $\vecy_\mathrm{V}$.
This optimization-validation process is done for $K$ combinations of $\vecy^\PHDG_\mathrm{T}$ and $\vecy^\PHDG_\mathrm{V}$.

There are two mean-squared errors (MSEs) that quantify the solution.
One is the training error defined by
\begin{align}
    \mathrm{MSE}_\mathrm{T} =
    \frac{1}{M_\mathrm{T}} \| \bm{y}^\PHDG_\mathrm{T} - \mathcal{P}^\PHDG_\mathrm{T} A \bm{x}^{\ast} \|_2^2,
\end{align}
where ${M_\mathrm{T}=M(K-1)/K}$ is the dimension of vector $\vecy^\PHDG_\mathrm{T}$, and $\mathcal{P}^\PHDG_\mathrm{T}$ is a projection operator onto the subspace that $\vecy^\PHDG_\mathrm{T}$ belongs to.
$\mathrm{MSE}_\mathrm{T}$ exhibits a monotonic growth with increasing $\lambda$ as shown in Fig.~\ref{fig:cv}(a), because $\lambda$ directly controls the ratio between $\mathrm{MSE}_\mathrm{T}$ to the $L_1$-norm regularization term [see Eq.~\eqref{eq:lasso-fourier}].

The second quantity is called the validation error or the CV error, which is defined by
\begin{align}
    \mathrm{MSE}_\mathrm{V} =
    \frac{1}{M_\mathrm{V}} \| \bm{y}^\PHDG_\mathrm{V} - \mathcal{P}^\PHDG_\mathrm{V} A \bm{x}^{\ast} \|_2^2,
\end{align}
where ${M_\mathrm{V}=M/K}$ and ${\mathcal{P}_\mathrm{V}=1-\mathcal{P}_\mathrm{T}}$.
$\mathrm{MSE}_\mathrm{V}$ represents to what extent the fitting result is general.
Here, ``general'' means the ability that the results infer different dataset.
If $\vecx^{\ast}$ is designed to fit minute structure due to noise in $\vecy^\PHDG_\mathrm{T}$, $\vecx^{\ast}$ would not match other dataset, namely, $\vecy^\PHDG_\mathrm{V}$.
The validation error thus gets worse as $\lambda$ decreases beyond a reasonable region.
We determined an optimal value of $\lambda$ by the minimum of MSE$_\mathrm{V}$, which yields ${\lambda=1.0 \times 10^{-1}} {\equiv \lambda_\mathrm{opt}}$ as indicated by the dashed vertical line in Fig.~\ref{fig:cv}(a).

The obtained optimal value is better understood by analyzing the effect of the $L_1$-norm regularization.
Figure~\ref{fig:cv}(b) shows the number of non-zero components $N_{>0}$ of the $L_1$ term [the second term in Eq.~\eqref{eq:lasso-fourier}].
Below the optimal $\lambda$, $N_{>0}$ rapidly increases as $\lambda$ decreases.
Such components that appear only for small-$\lambda$ are used to fit minute structure of the input data, that is, noise, and thus increase validation errors.
At ${\lambda=\lambda_\mathrm{opt}}$, $2{,}103$ components are finite out of ${3N_\mathrm{irr}=119{,}133}$ in $D_{\xi} \vecrho$, meaning that only $1.8\%$ ($98.2\%$) of $\nabla \rho(\vecp)$ are finite (zero) in the final result.

\section{Noise-Level Dependence}
\label{sec:results}

\begin{figure*}
    \centering
    \includegraphics[width=0.98\linewidth]{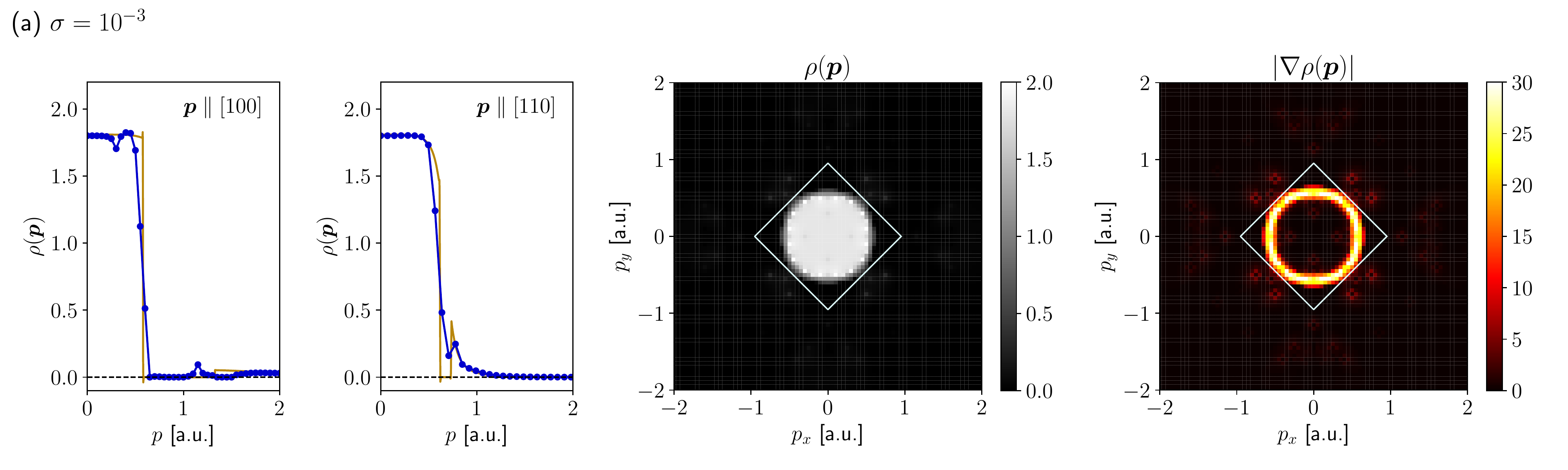}
    \includegraphics[width=0.98\linewidth]{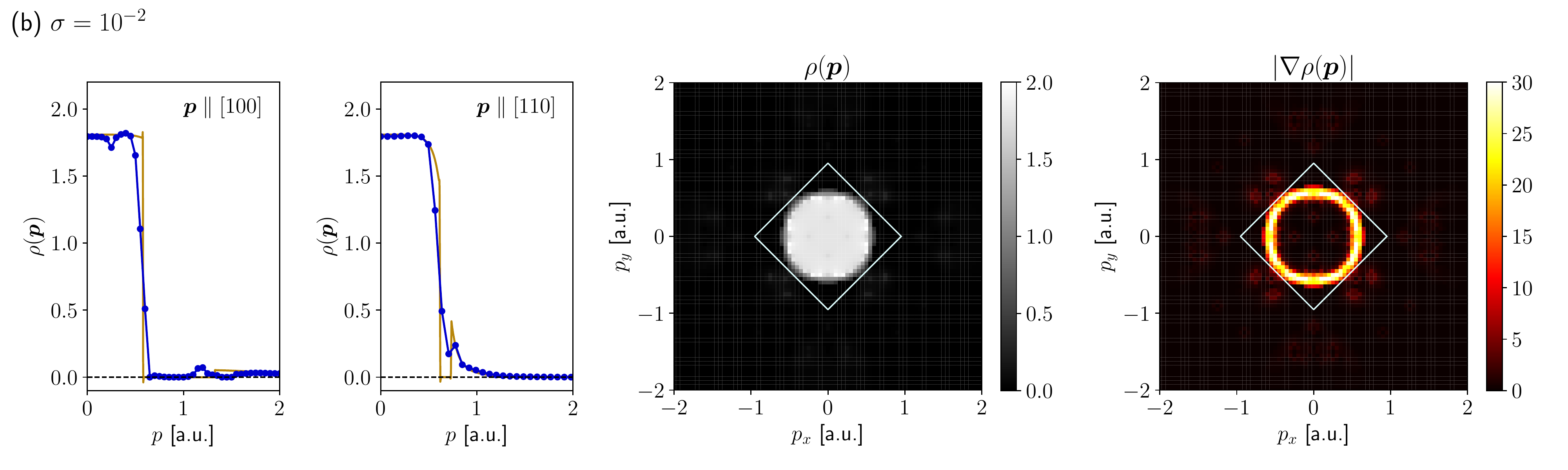}
    \includegraphics[width=0.98\linewidth]{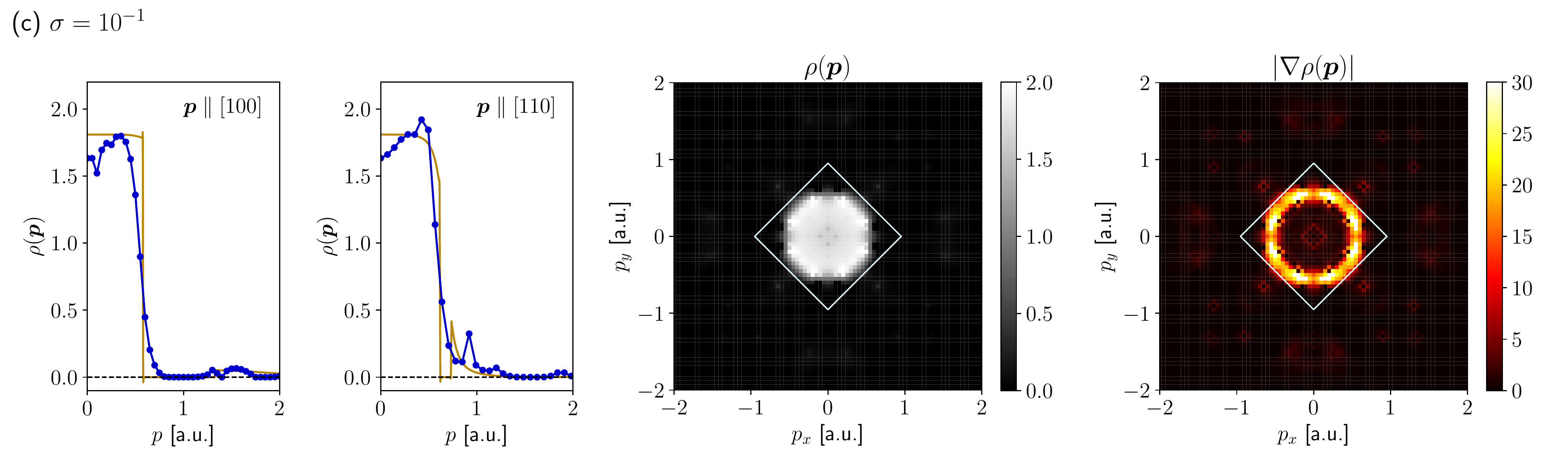}
    \caption{The reconstructed EMD results for different noise levels, (a) ${\sigma=10^{-3}}$, (b) $10^{-2}$, (c) $10^{-1}$. The panels from left to right present $\rho(\vecp)$ for ${\vecp \parallel [100]}$,  $\rho(\vecp)$ for ${\vecp \parallel [110]}$, intensity maps of $\rho(\vecp)$ on the ${p_z=0}$ plane, and $| \nabla \rho(\vecp)|$ on the ${p_z=0}$ plane, respectively.
    The square in the intensity maps indicates the first Brillouin zone.}
    \label{fig:np_multi}
\end{figure*}

This section focuses on the influence of noise on the reconstructed EMD results.
Fig.~\ref{fig:np_multi} compares the reconstructed EMD for different noise levels, ${\sigma=10^{-3}}$, $10^{-2}$, and $10^{-1}$.
The regularization parameter $\lambda$ were optimized separately using the CV method.
The left two panels in Fig.~\ref{fig:np_multi}(a) are replots of \ref{fig:np_example}(b) in a different range. The third panel shows the intensity map of $\rho(\vecp)$ on the ${p_z=0}$ plane.
It is clear that the occupied states are almost isotropic and resembles those of free-electron gas.
The Fermi surface can be emphasized by taking the gradient of $\rho(\vecp)$.
The intensity map of $|\nabla \rho(\vecp)|$ is presented in the right-most panel.
The high-intensity circle indicates the Fermi surface, which could be compared with the ARPES spectrum at zero frequency.

As the noise level increases from ${\sigma=10^{-3}}$ to ${\sigma=10^{-1}}$ [Fig.~\ref{fig:np_multi}(a) to Fig.~\ref{fig:np_multi}(c)], the discontinuity in $\rho(\vecp)$ gets blurred. Correspondingly, the peak in $|\nabla \rho(\vecp)|$ becomes broader.
These results demonstrate that the noise level affects the momentum resolution in the reconstructed $\rho(\vecp)$.
Nevertheless, we can still determine the Fermi surface by tracking the ridge in $|\nabla \rho(\vecp)|$.

\section{Summary}
\label{sec:summary}

The inverse problem for reconstruction of the three-dimensional EMD, $\rho(\vecp)$, is underdetermined in nature, because the number of experimentally measured scattering directions is limited.
We employed the compressed sensing, which can deal precisely with undertermined systems.
The compressed sensing uses the limited information for determination of EMD in a specific regions, that is, around the Fermi surface.
This is accomplished by the sparsity condition for $\nabla \rho(\vecp)$ implemented as an optimization problem called the generalized LASSO.

We tested this new technique on the reconstruction of $\rho(\vecp)$ of bcc-Li from the Compton profiles computed by DFT.
The compressed-sensing technique allows us to reconstruct $\rho(\vecp)$ from 14 projections and to characterize the shape of the Fermi surface.
We also investigated the noise dependency in the reconstruction problem.
We show that even if the Compton profiles are perturbed by the noise (assumed errors of experimental measurements), 
our method stably captures the feature around the Fermi surface.

The demonstration with bcc-Li will lead to further applications to more complicated materials.
We believe that out method based on the compressed sensing will contribute to accelerate the research into fermiology and stimulate development on the experiment side as well.

\appendix

\begin{acknowledgments}
This work was supported by JSPS KAKENHI grants No. 17K12749, No. 19K03649, No. 20K20522, No. 21H01003, and No. 21H01041.
MM was supported by JST CREST (JPMJCR1861).
LC acknowledges the financial support by the Deutsche Forschungsgemeinschaft through TRR80 (project E2) Project number 107745057.
\end{acknowledgments}

\section{Discrete cosine transformation}
\label{app:fourier}

The Fourier transform in Eq.~\eqref{eq:B_z} is computed as follows.
Let us assume that we have $N$ data of $J(p_z)$ on a uniform grid in the range ${p_z=[0:p_\mathrm{max}]}$ with the step size ${\Delta p_\mathrm{exp} = p_\mathrm{max}/(N-1)}$.
The data set is represented by $p_k$ and $J_k$ with $k=0, 1, \cdots, N-1$.
Discretizing the integral in Eq.~\eqref{eq:B_z} between $-p_\mathrm{max}$ and $p_\mathrm{max}$ and using the relation ${J(p_z)=J(-p_z)}$, we obtain
\begin{align}
    B_n = \Delta p_\mathrm{exp} \left[  J_0 + (-1)^n J_{N-1} + 2 \sum_{k=1}^{N-2} J_k \cos \left( \frac{\pi kn}{N-1} \right) \right],
\end{align}
where $B_n$ is defined by ${B_n \equiv B(0, 0, z_n)}$ with $z_n$ being ${z_n \equiv n\pi / p_\mathrm{max}}$ ($n=0, 1, \cdots, N-1$).
This discrete cosine transformation is classified as Type I in SciPy python package.

\section{Symmetry}
\label{app:symmetry}
Symmetry of $\rho(\vecp)$ in the momentum space plays crucial roles in reducing the number of $\vecp$-points to save memory and improving the accuracy of the reconstruction.

We first make a grid in the whole three-dimensional space.
In the Cartesian coordinate, the momenta $p_x$, $p_y$, and $p_z$ are discretized into $L$ points severally in the range ${[-P_\mathrm{max}, P_\mathrm{max}]}$.
We thus obtain ${N=L^3}$ grid points, which are represented by $\vecp_j$.

We transform the vector $\vecp_j$ into ${\vecp_j'=\mathcal{R} \vecp_j}$ by symmetry operations $\mathcal{R}$ that are invariant in the crystal.
In the case of O$_h$ point-group symmetry, there are 48 operations.
If $\vecp'_j$ corresponds to a grid point, say $\vecp_k$, we regard that two vectors $\vecp'_j$ and $\vecp_k$ are equivalent. Applying all symmetry operations $\{ \mathcal{R} \}$ to all grid points $\{ \vecp_j \}$, we construct an inequivalent set of vectors, which we represent by $\{ \tilde{\vecp}_j \}$.

Typical choices of $L$ are summarized in Table~\ref{tab:size} together with the corresponding values of ${N=L^3}$ and the number of inequivalent vectors, $N_\mathrm{irr}$.
As expected, we obtain ${N_\mathrm{irr}/N \sim 1/48}$.
In the case with the ADMM algorithm, whose memory and computation cost scales $O(N^3)$, we can deal with up to ${N_\mathrm{irr}\sim 10^4}$ with desktop computers and $10^5$ with cluster computers, namely, ${L \simeq 81}$ and $161$, respectively.

\begin{table}[]
    \centering
    \begin{tabular}{ccc}
    \hline
        $L$ & $N$ & $N_\mathrm{irr}$  \\
        \hline
        41 & 68,921 & 1,771 \\
        61 & 226,981 & 5,456 \\
        81 & 531,441 & 12,341 \\
        101 & 1,030,301 & 23,426 \\
        121 & 1,771,561 & 39,711 \\
        141 & 2,803,221 & 62,196 \\
        161 & 4,173,281 & 91,881 \\
        181 & 5,929,741 & $\sim 1.2 \times 10^5$ \\
        201 & 8,120,601 & $\sim 1.7 \times 10^5$ \\
        \hline
    \end{tabular}
    \caption{The number $N$ of $\vecp$-grid points for representing $\rho(\vecp)$ and the number $N_\mathrm{irr}$ in the irreducible region.}
    \label{tab:size}
\end{table}

The symmetry property is integrated into computations to have only $N_\mathrm{irr}$ instead of $N$ as follows.
We introduce notations $\tilde{\vecrho}$ for the set of the momentum density at the inequivalent points, and $\vecrho$ for the full set of the momentum density at the whole points. 
By definition, $\vecrho$ is obtained by upfolding $\tilde{\vecrho}$ by
\begin{align}
    \vecrho = F \tilde{\vecrho},
    \label{eq:upfold}
\end{align}
where $F$ is ${(N\times N_\mathrm{irr})}$ matrix which has one 1 in each row and 0 otherwise.
The matrix-vector product $A \vecrho$ is then evaluated as
\begin{align}
    A \vecrho = A F \tilde{\vecrho} \equiv \tilde{A} \tilde{\vecrho},
    \label{eq:downfold}
\end{align}
where $\tilde{A}$ is a matrix that is downfolded from $A$ by ${\tilde{A} \equiv A F}$.
The size of the original matrix $A$ is ${(M \times N)}$, while $\tilde{A}$ is ${(M \times N_\mathrm{irr})}$.
Using $\tilde{A}$, an actual evaluation of the $L_2$ term, ${\| \vecy - A \vecrho \|_2^2}$, is done with ${\| \vecy - \tilde{A} \tilde{\vecrho} \|_2^2}$, and the solution for $\tilde{\vecrho}$ is evaluated.
Finally, $\tilde{\vecrho}$ is upfolded into $\vecrho$ using Eq.~\eqref{eq:upfold}.

The evaluation of $L_1$ term, $\| D \vecrho \|_1^\PHDG$, needs further elaborate treatment, because the above downfolding reduces ${(3N \times N)}$ matrix $D$ to ${(3N \times N_\mathrm{irr})}$ matrix $(DF)$, which still has the scale $N$.
In order to eliminate the $N$-scale in $\| D \vecrho \|_1^\PHDG$, we remark that $D \vecrho \equiv \vecrho'$ has the same symmetry property as $\vecrho$, since $D$ represents the derivative, which preserves symmetry.
We can therefore sum up over equivalent elements in $\vecrho'$ before its $L_1$ norm $\| \vecrho' \|_1^\PHDG$ is evaluated.
This leads the equality ${\| D\vecrho \|_1^\PHDG = \| F^\mathrm{T} (D\vecrho) \|_1^\PHDG}$.
Substituting $\vecrho$ with $\tilde{\vecrho}$ using Eq.~\eqref{eq:upfold}, we obtain
\begin{align}
    \| D\vecrho \|_1^\PHDG = \| F^\mathrm{T} D F \tilde{\vecrho} \|_1^\PHDG \equiv \| \tilde{D} \tilde{\vecrho} \|_1^\PHDG,
\end{align}
where the matrix $\tilde{D}$ is defined by ${\tilde{D} \equiv F^\mathrm{T} D F}$.
The size of $\tilde{D}$ is ${(3N_\mathrm{irr} \times N_\mathrm{irr})}$, and thus the scale $N$ has been eliminated.

\section{ADMM for generalized LASSO with constraints}
\label{app:admm}
We consider a generalized LASSO problem with additional constraints.
The function to minimize is $\mathcal{F}(\vecx)$ in Eq.~\eqref{eq:genlasso}.
Two constraints, non-negativity and a sum-rule, are generalized into
\begin{align}
    P\vecx \geq 0,
    \quad
    \langle S \vecx \rangle = s,
    \label{eq:constraint}
\end{align}
where the bracket stands for $\langle S\vecx \rangle \equiv \sum_j (S\vecx)_j$, and $s$ is a constant.
The matrices, $A$, $B$, $P$, and $S$, have the same column size $N$, but their row sizes are, in general, all different.

We solve this optimization problem using ADMM by Boyd \textit{et al.}~\cite{Boyd2010}.
A situation similar to the present case with constraints is considered in Refs.~\cite{Otsuki2017,Otsuki2020}.
Here, we generalize them to includes general four matrices $A$, $B$, $P$, and $S$.

Introducing auxiliary vectors $\vecz$ and $\vecz'$, we rewrite the function $\mathcal{F}$ in Eq.~\eqref{eq:genlasso} as
\begin{align}
  \begin{split}
    \widetilde{\mathcal{F}}(\vecx, \vecz, \vecz')
    &= \frac{1}{2} \| \vecy -A\vecx \|_2^2
    -\nu (\langle S\vecx \rangle -s)
    \\
    &+ \lambda \| \vecz \|_1^\PHDG
    + \lim_{\gamma\to\infty} \gamma \sum_j \Theta(-z'_j),
  \end{split}
    \label{eq:genlasso_admm}
\end{align}
where $\nu$ is a Lagrange multiplier that enforces the sum-rule constraint.
With the conditions
\begin{align}
    \vecz = B \vecx, \quad \vecz' = P \vecx,
    \label{eq:constraint_admm}
\end{align}
Eq.~\eqref{eq:genlasso_admm} is reduced to Eq.~\eqref{eq:genlasso} plus the constraints in Eq.~\eqref{eq:constraint}.
The advantage of the latter form, $\widetilde{\mathcal{F}}(\vecx, \vecz, \vecz')$, is that the minimization with respect to $\vecx$, $\vecz$, and $\vecz'$ can be done using analytical formulas.
Therefore, our task is to make $\vecx$, $\vecz$, and $\vecz'$ satisfy the condition, Eq.~\eqref{eq:constraint_admm}, keeping minimizing $\widetilde{\mathcal{F}}$.

In the ADMM approach, the constraints, Eq.~\eqref{eq:constraint_admm}, is imposed by the augmented Lagrange multiplier method.
We here quote the update formulas from Ref.~\cite{Otsuki2020} with generalization to the four-matrices representation:
\begin{align}
    \vecx &\leftarrow \left( A^\mathrm{T} A + \mu B^\mathrm{T} B + \mu' P^\mathrm{T} P \right)^{-1} 
    \nonumber \\
    &\quad \times \left( A^\mathrm{T} \vecy + \mu B^\mathrm{T} (\vecz - \vecu) + \mu' P^\mathrm{T} (\vecz' - \vecu') + \nu S^\mathrm{T} \bm{d} \right)
    \nonumber \\
    &\quad \equiv \bm{\xi}_1 + \nu \bm{\xi}_2,
    \label{eq:admm_update1}
    \\
    \vecz &\leftarrow \mathcal{S}_{\lambda/\mu} (B\vecx + \vecu),
    \\
    \vecu &\leftarrow \vecu + B\vecx - \vecz,
    \\
    \vecz' &\leftarrow \mathcal{P}_{+} (P\vecx + \vecu'),
    \\
    \vecu' &\leftarrow \vecu' + P\vecx - \vecz',
\end{align}
where $\bm{d}$ is a vector with all elements being 1, 
$\mathcal{P}_+$ is define by ${\mathcal{P}_+(x)=\max (x, 0)}$,
which truncates negative values to zero,
and $\mathcal{S}$ is the element-wise soft-thresholding function, which is defined for each element by
\begin{align}
    \mathcal{S}_{\lambda} (x) =
    \begin{cases}
    0 & (|x| \leq \lambda) \\
    x - \mathrm{sgn}(x) \lambda & (|x| > \lambda)
    \end{cases}.
\end{align}
The Lagrange multiplier $\nu$ is determined by
\begin{align}
    \nu = \frac{s-\langle S \bm{\xi}_1 \rangle}{\langle S \bm{\xi}_2 \rangle}.
\end{align}
The parameter $\mu$ and $\mu'$ are penalty parameters,
which will be explained later.
As an initial condition, all vectors $\vecx$, $\vecz$, $\vecu$, $\vecz'$, and $\vecu'$ are set to zero.

The most expensive computation in this calculation is the matrix inversion in Eq.~\eqref{eq:admm_update1}.
We compute the LU decomposition of the matrix $M \equiv A^\mathrm{T} A + \mu B^\mathrm{T} B + \mu' P^\mathrm{T} P$ before starting the iteration~\footnote{The Cholesky decomposition can be applied instead of the LU decomposition, because the matrix $M$ is real symmetric. However, we confirmed that the LU decomposition was faster in our implementation using SciPy.}.
Using this result, linear equations are solved in each iteration to update $\vecx$.
The cost for the LU decomposition is $O(N^3)$, while the cost for the updates is $O(N^2)$, where $N$ is the dimension of $\vecx$ ($N$ should be replaced with $N_\mathrm{irr}$ when the symmetry is applied as presented in Appendix~\ref{app:symmetry}).
Therefore, the computational cost and the memory storage required for the LU decomposition determine the upper limit of the system size.

Convergence of the iteration should be checked in two perspectives.
One is the residual error of the constraint~(\ref{eq:constraint_admm}), namely, ${r\equiv \| \vecz - B \vecx \|_2^\PHDG}$.
The other is convergence of the variables, e.g., ${s\equiv \| \vecz_{k+1} - \vecz_k \|_2^\PHDG}$, where $k$ indicates the quantity at the $k$-th iteration.
A fast convergence is achieved when $r$ and $s$ are of the same order.
A relative magnitude between $r$ and $s$ depends on $\mu$:
Larger values of $\mu$ reduce $r$, since $\mu$ is the penalty against the constraints~(\ref{eq:constraint_admm}).
Therefore, if convergence of $r$ is slower than $s$, one should increase $\mu$, and vise versa.
See Ref.~\cite{Boyd2010} for more details.

\bibliography{compton, sparse_modeling}

\end{document}